\documentclass[letterpaper,runningheads]{llncs}
\usepackage[utf8]{inputenc}
\usepackage[T1]{fontenc}
\usepackage{graphicx}
\usepackage[hidelinks]{hyperref}
\usepackage{svg}
\usepackage{amssymb,amsmath} 
\usepackage[section]{placeins}
\usepackage{blindtext}
\usepackage{enumitem}
\usepackage{multirow}
\usepackage{booktabs}
\usepackage{footmisc}
\usepackage{cite}
\usepackage{pdfpages}
\usepackage{wrapfig}
\usepackage{tikz}
\graphicspath{{}}
\DeclareGraphicsExtensions{.pdf,.jpg,.png}

\usepackage{mdwtab}
\usepackage{url}
\usepackage{array}
\usepackage{ragged2e}
\newcolumntype{P}[1]{>{\RaggedRight\hspace{0pt}}p{#1}}
\begin{document}
\title{Modeling and Analysis of \\ Boundary Objects and Methodological Islands \\in Large-Scale Systems Development}
\titlerunning{Modeling and Analysis of Boundary Objects and Methodological Islands}

\author{Rebekka Wohlrab\inst{1,2}\orcidID{0000-0002-5449-7900} \and Jennifer Horkoff\inst{1}\orcidID{0000-0002-2019-5277} \and Rashidah Kasauli\inst{1}\orcidID{0000-0003-1655-4377} \and Salome Maro\inst{1}\orcidID{0000-0003-1560-6833} \and Jan-Philipp Stegh\"ofer\inst{1}\orcidID{0000-0003-1694-0972} \and Eric Knauss\inst{1}\orcidID{0000-0002-6631-872X}}
\authorrunning{R. Wohlrab et al.}

\institute{\mbox{{Chalmers $|$ University of Gothenburg, Sweden \and Systemite AB, Sweden}}\\
	\email{\{wohlrab, jenho, rashida\}@chalmers.se, \{salome.maro, jan-philipp.steghofer, eric.knauss\}@cse.gu.se}}

\maketitle

\begin{abstract}
	\begin{tikzpicture}[overlay]
	\node[draw, fill=white, thick, align=center, style={inner sep = 4}] (b) at (4,8){This is the author version of our paper accepted to the 39th International Conference on Conceptual Modeling (ER 2020).\\	
		The final authenticated version will be available online in the Lecture Notes in Computer Science by Springer.};
	\end{tikzpicture}
	Large-scale 
	{systems development} commonly face
	{s} the challenge of managing relevant knowledge between different organizational groups, particularly in increasingly agile contexts.
	In previous studies, we found the importance of analyzing methodological islands (i.e., groups using different development methods than the surrounding organization) and boundary objects between them.
	In this paper, we propose a metamodel to better capture and analyze coordination and knowledge management in practice.
	Such a metamodel can allow practitioners to describe current practices, analyze issues, and design better-suited coordination mechanisms.
	We evaluated the conceptual model together with four large-scale companies developing complex systems.  
	In particular, we derived an initial list of bad smells that can be leveraged to detect issues and devise suitable improvement strategies for inter-team coordination in large-scale development.
	We present the model, smells, and our evaluation results.\newline
	
	\textbf{Keywords:} {boundary objects $\cdot$ agile development $\cdot$ empirical studies.}
\end{abstract}

\section{Introduction}
\label{intro}
Large-scale systems engineering companies commonly face the challenge of coordination between multiple and multidisciplinary teams (e.g., software, systems, hardware).
Especially in large-scale agile development, inter-team coordination is a recognized challenge~\cite{Dingsoyr2017}.
In practice, ways of working are not universal in large companies. 
Teams are surrounded by other organizational parts that do not use the same methods---and thus become ``\textit{methodological islands}''~\cite{Kasauli2020ICSSP}.
For instance, in a large automotive company, more than 500 teams exist, using diverse practices (agile, waterfall), with complex interdependencies and multiple suppliers.
Coordination is supported by various artifacts (e.g., written documents, models, backlogs, or code).
Furthermore, phone calls, meetings in communities of practice, and other mechanisms are used to coordinate concerns around these artifacts.
In such a situation, it can be challenging to coordinate knowledge between different organizational groupings.
Practitioners need to better understand the factors causing these groups (or islands) to cluster or form and the effectiveness of the current ways of supporting communication.
For example, is a particular written document between two islands fit for coordination?  Is it too flexible or too rigid?  Is it both complex and changing frequently? Is it governed, and do those that govern the document understand its use? 
Can the current coordination situation be understood, made explicit, and improved?

In previous studies, we have aimed to characterize these coordination needs by focusing on methodological islands (MIs) and boundary objects (BOs)~\cite{Kasauli2020ICSSP}.
Boundary objects create a common understanding between groups and can facilitate inter-team coordination and knowledge management~\cite{Wohlrab2019JSME}.
We have investigated the nature and use of these BOs in practice, but we have not yet created a method to systematically capture BOs and MIs (BOMIs) in a structured way. 
To the best of our knowledge, there is no modeling approach and conceptual model available to specifically address boundary objects and methodological islands.  

In this paper, we address this gap by proposing a metamodel for boundary objects and methodological islands in large-scale systems development.
This model is based on empirical data and accounts from ongoing projects~\cite{Kasauli2020ICSSP,Wohlrab2020JSS,Wohlrab2019ICSASurvey}.
By creating such a model, a complete picture of an organization's coordination needs and boundary objects can be established, analyzed, and used to identify and mitigate current issues in a more visual and structured way.

We evaluated the metamodel together with four large-scale systems companies and describe the corresponding instance models created.
We present initial findings on how the model can be used to identify bad smells and issues.

This paper is organized as follows: Sec.~\ref{sec:Background} presents the background. Sec.~\ref{sec:findings} describes our metamodel, method and smell description, followed by the evaluation in Sec.~\ref{Evaluation}. Sec.~\ref{sec:realted} briefly reviews related modeling approaches,   Sec.~\ref{Discussion} discusses our findings and describes threats to validity, while Sec.~\ref{Conclusions} concludes the paper.

\section{Background}
\label{sec:Background}
We describe background information to motivate this paper's contributions.

\textbf{Boundary Objects.} 
Boundary objects (BOs) are ``\textit{objects which are both plastic enough to adapt to local needs and the constraints of the several parties employing them, yet robust enough to maintain a common identity across sites}''~\cite{Star1989}.
The concept was initially coined in sociology and has proven to be useful in a variety of domains.
Recently, BOs have increasingly been studied in software and systems engineering~\cite{Sedano2019,Zaitsev2016,Wohlrab2019JSME}.

Over the last two years, we have engaged with four large-scale systems engineering companies to support them in adopting agile methods and managing important knowledge.
We used the design science methodology~\cite{Hevner2004} to investigate coordination in large-scale systems engineering, develop suitable \emph{design artifacts} targeting practical problems, and evaluate them in several iterations.
We build upon the findings of this long-term project.
In Sec.~\ref{Evaluation}, we describe the participating companies in further detail.

As part of our work on BOs, we conducted several studies.
We analyzed currently used artifacts and created guidelines to manage them in large-scale agile contexts, including concerns related to the level of detail and versioning of these artifacts~\cite{Wohlrab2019JSME}.
We found that BOs can belong to several super types (e.g., Technology, Task, or Planning)~\cite{Kasauli2020ICSSP} and should be managed in groups of representatives of several teams~\cite{Wohlrab2019JSME}.
Moreover, we studied architecture descriptions and interfaces as BOs~\cite{Wohlrab2019ICSASurvey, Wohlrab2019ICSAInterface}.
We found that important dimensions of interface change are stability, time to perform a change, criticality, level of abstraction, distance to affected parties, number of affected components, position in the interface's lifecycle, and maturity of affected functions.
Moreover, many companies describe \textit{information models} to capture artifact types and their relations.
These information models also serve as BOs, change over time, and can be used to define the required degree of alignment of different teams' practices~\cite{Wohlrab2020JSS}.

BOs are commonly used between individuals from several (sub-)disciplines, who refer to concepts with different terminologies~\cite{Wohlrab2019JSME}.
The groups using BOs need to be properly understood to enable inter-team coordination.

\textbf{Methodological Islands.} 
The mix of methods in large-scale organizations is a recognized challenge~\cite{Wohlrab2019JSME}.
In our empirical study on large-scale development, agile teams were described as ``\textit{agile islands in a waterfall}''~\cite{Kasauli2020ICSSP}.
This phenomenon is not limited to the discrepancy of agile and plan-driven methods, but a general issue.
Therefore, we use the term \textit{methodological islands} (MIs) for organizational groups using different development methods than the surrounding organization.
We identified that MIs can be of different types, e.g., individual teams (e.g., component teams), groups of teams (e.g., departments), or entire organizations.
MIs arise due to several \textit{drivers} related to \textit{business}, \textit{process}, and \textit{technology}.

Based on these studies, we got an understanding of BOs and MIs in large-scale systems engineering.
These findings needed to be better instrumentalized to support practitioners, in particular, using a systematic approach to capture BOs and MIs~\cite{Kasauli2020ICSSP}.
Such an approach would constitute a formal treatment to describe and evaluate coordination needs.

\section{BOMI Metamodel, Method, and Analysis}
\label{sec:findings}
\begin{figure}
	\centering
	\includegraphics[width=\linewidth]{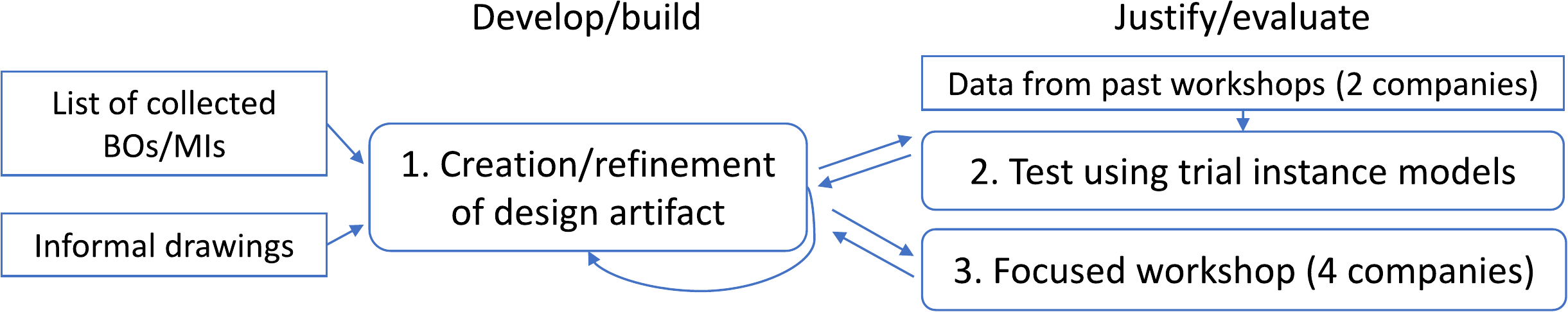}
	\caption{Overview of steps of our research method}
	\label{fig:methodoverview}
	\vspace{-0.2cm}
\end{figure}

In the following, we present our main contributions, i.e., the BOMI metamodel, method, and analysis capabilities provided by the model.
We continued our design science approach~\cite{Hevner2004} but with a focus on developing a metamodel, modeling guidelines, and model smells.
An overview of the input artifacts and steps of our method is shown in Figure~\ref{fig:methodoverview}.
We went through several iterations designing an artifact (metamodel, method, and smells) and performing evaluations of the artifact both locally and with four companies.
In the first round, we used our informal drawings and lists of collected BO and MIs in practice, along with our knowledge gathered from the companies, to come up with a first draft of the artifact.
The paper authors discussed the artifact and made local improvements.
We then used historical data gathered from workshops with two of the companies to create trial models of their BOMI situation.
After discussion, this caused further iteration over the artifact.
Finally, we evaluated the design artifact in a focused workshop (see Sec.~\ref{Evaluation}).
The four companies we collaborated with are described in Table~\ref{tab:participants}.
\begin{table*}[b]
	\centering
	\caption{Descriptions of participating companies.}
	\label{tab:participants}
	\begin{tabular}{@{}p{.13\textwidth}p{.855\textwidth}@{}}
		\toprule
		Company A & Develops telecommunications products.
		Separate organizational units exist for sales, product management, and other purposes. \\
		Company B & Develops mechanical products, both for consumer markets and for industrial development and manufacturing. The systems are decomposed into several elements, which is also reflected in the organizational structure. \\
		Company C & Is an automotive Original Equipment Manufacturer (OEM). 
		Traditionally, the company has been structured according to vehicle parts (e.g., powertrain, chassis, ...), but has undergone restructuring into agile teams. \\
		Company D & Develops high-tech solutions for vehicular systems.
		Software development teams are largely independent of hardware development.
		\\
		\bottomrule
	\end{tabular}
\end{table*}
\subsection{BOMI Metamodel}

\begin{figure}
	\centering
	\includegraphics[width=0.99\linewidth]{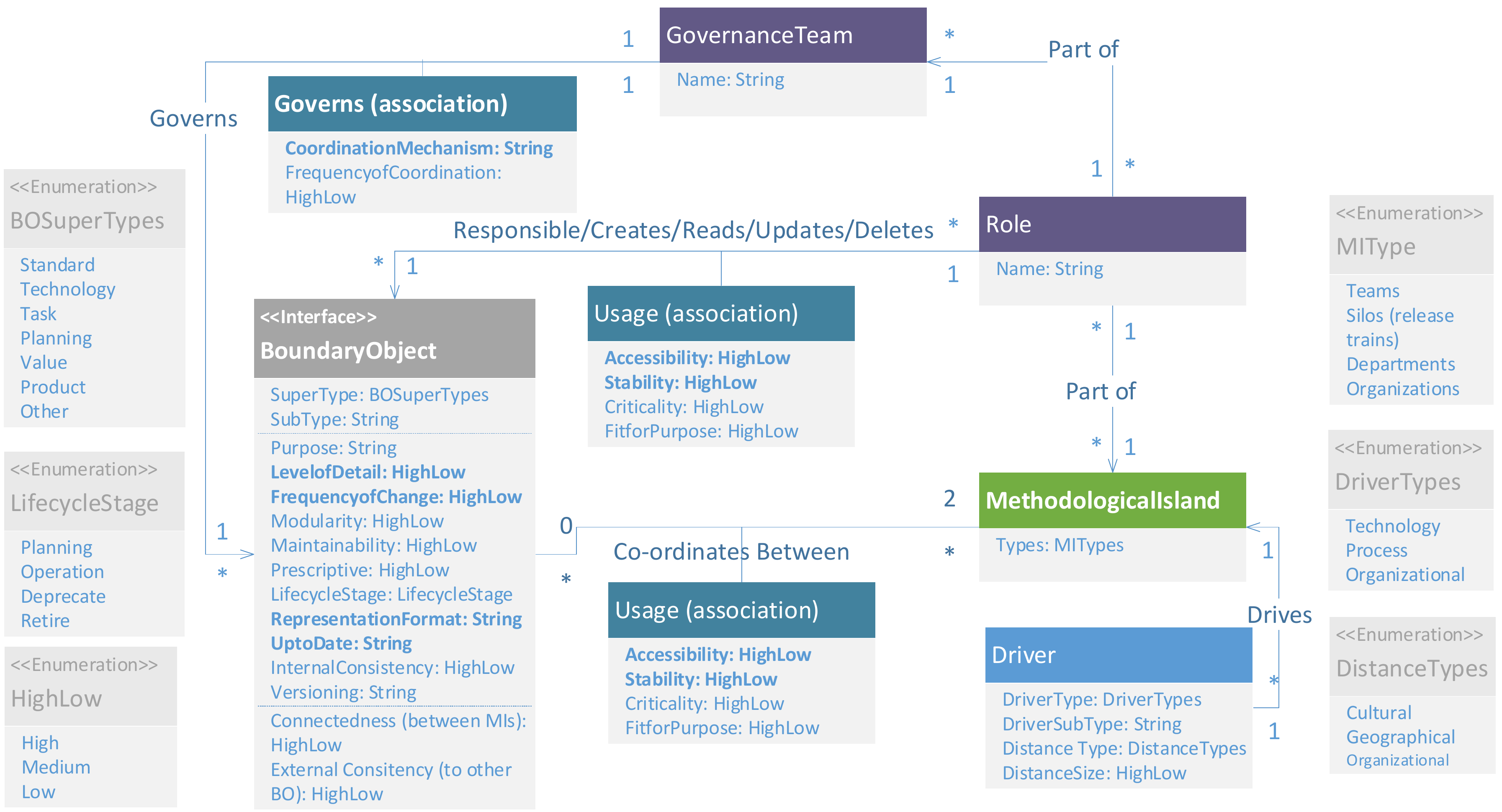}
	\caption{Metamodel for Boundary Objects and Methodological Islands (BOMI)}
	\label{fig:conceptualmodel}
\end{figure}

To capture our conceptual model, we use a UML class diagram.
Other languages could work just as well, but we choose UML due to its familiarity.
The latest version of the BOMI metamodel can be found in Fig.~\ref{fig:conceptualmodel}. 

Based on our past findings, the most critical element of the metamodel is the \texttt{BO} itself (in dark gray).
We label this class as an interface, given the nature of BOs as interfaces between methodological islands.
We give this class a \textsl{SuperType} and \textsl{SubType}, based on our past classification findings~\cite{Kasauli2020ICSSP}.
The \textsl{SuperType} is an enumeration, with a set list of options, while we found an enumeration was too restrictive for the \textsl{SubType}, and leave this as free text (a String).

We use our experiences to identify a number of internal \texttt{BO} attributes, including the \textsl{Purpose}, \textsl{Level of detail}, \textsl{Frequency of change}, \textsl{Level of modularity} and \textsl{Maintainability}, whether the \texttt{BO} represents \textsl{Prescriptive} knowledge (as opposed to descriptive), which \textsl{Lifecycle} stage the \texttt{BO} is used in, with an enumeration of four options (\textsl{Planning}, \textsl{Operation}, \textsl{Deprecate}, \textsl{Retire}), \textsl{Representation Format} (e.g., free text, model, table), the level of \textsl{Internal Consistency}, and what sort of \textsl{Versioning} information it may have.
The last two attributes describe the relationship between this \texttt{BO} and other classes in the model, in this case,  \textsl{Connectedness} of the \texttt{MIs} using the \texttt{BO}, and how  \textsl{Externally Consistent} it is with other \texttt{BO} instances.
These attributes are either free text (String) or are described via a simple qualitative scale of  \textsl{High},  \textsl{Medium}, and  \textsl{Low} (the \texttt{HighLow} enumeration).
We found that although this qualitative scale can be used for a quick summary, often a more complex description is needed.
For example, for architecture descriptions, the level of detail of the \texttt{BO} changes depending on the  \textsl{Lifecyle} stage.
Thus, we find the need to accompany each attribute with a short explanation of the value.
We omit this from the current metamodel for simplicity, but note that the instance models should be accompanied by some explanatory text.

A \texttt{Methodological Island} (in green) contains an enumeration of types based on our past findings (\textsl{Teams}, \textsl{Silos}, \textsl{Departments}, \textsl{Organizations}).
For \texttt{MIs}, the relations to other elements are crucial.
Organizational \texttt{Roles}, with role \textsl{names} are part of the \texttt{MIs}.
A \texttt{Role} is responsible for, or has a CRUD relationship with a \texttt{BO}.
The \texttt{Usage} association class between these classes captures how \texttt{Roles} use \texttt{BOs}.
We can model a \texttt{BO}'s accessibility for a \texttt{Role}, its \textsl{Stability}, \textsl{Criticality}, and whether it is \textsl{Fit for Purpose}.
Ideally, a \texttt{Role} is part of a \texttt{MI}, and the \texttt{Role}'s interaction with the \texttt{BO} is described in the \texttt{Usage} class.
In some cases, practitioners were reluctant to explicitly model roles and only model \texttt{BOs} and \texttt{MIs}, either because the inclusion of \texttt{Roles} caused the model to drastically increase in size or because the \texttt{Role} and \texttt{MI} were similar (e.g., ``Development Team'' $\rightarrow$ Role should be ``Developer'').
Thus, we repeat this association class in two places and one can also create a \texttt{Usage} association between \texttt{BOs} and \texttt{MIs}.

Our past work uncovered the concept of MI drivers, the reason for the MI divide.
We capture that a \texttt{Driver} drives an \texttt{MI}, and describe possibly interesting attributes of the drivers, including an enumerated \textsl{Driver Type} (\textsl{Technology}, \textsl{Process}, \textsl{Organization}), a free-text \textsl{Driver SubType}, the \textsl{Distance Type}  culture/geography/organization inspired by~\cite{Bjarnason2017,holmstrom2006agile}, and the size of the distance.

Finally, based on past work~\cite{Wohlrab2019JSME,Wohlrab2019ICSASurvey}, we find that governance of BOs is crucial.
\texttt{Roles} can be part of a \texttt{Governance Team}, which governs a \texttt{BO}.
For instance, a Community of Practice is a potential governance team for architecture descriptions~\cite{Wohlrab2019ICSASurvey}.
We collect interesting attributes of this relationship in the \texttt{Governs} association class, including the \textsl{Coordination Mechanism} (e.g., meetings, processes, standards, tools), and the \textsl{Frequency of Coordination}.

Although other details could be added to this model, we aim for relative simplicity to better enable instantiation with and by our industrial partners. 

\subsection{BOMI Method}
\label{sec:method}
As part of our modeling workshops, we created a simple list of guiding questions based on our metamodel concepts and attributes, e.g., ``Which BO would you like to focus on?'', ``What roles interact with the BO?'', and ``Which islands do the roles belong to?''.
The full list of questions can be found in our online appendix\footnote[1]{\label{workshopmaterial}\url{https://doi.org/10.6084/m9.figshare.12363764.v1}}.
These questions are intended to guide in the creation of a BOMI instance model, either led by a modeling facilitator, or independently in a company.

\subsection{Instance Example}
\begin{figure}
	\centering
	\includegraphics[width=0.99\linewidth]{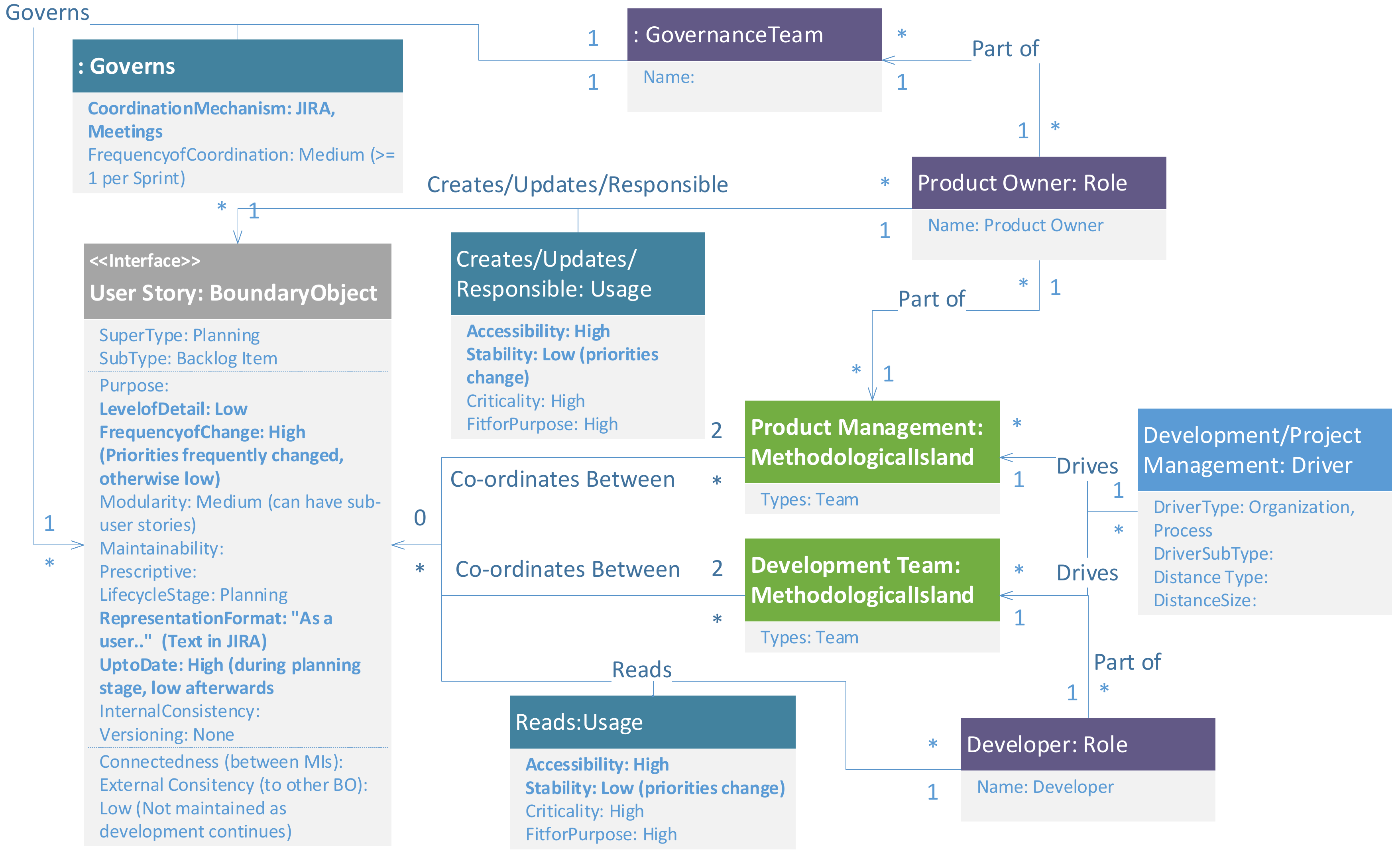}
	\caption{Instance model of BOMI setup for User Stories for Company A}
	\label{fig:Example1}
\end{figure}

To illustrate our model in action, we present an example derived from a workshop with our industrial partners in Fig.~\ref{fig:Example1}. More details about how this example was derived are provided in Sec.~\ref{Evaluation}.
For this example, we again use UML syntax.
In developing a BOMI language, we could create a domain-specific visual language, using customized icons or different shapes.
Although promising, we leave the exploration of a BOMI-specific visual syntax to future work, and instead use the visual syntax of UML, with the benefit of familiarity for our industrial partners. 

In this example, Company A (more detail in Sec.~\ref{Evaluation}) chose to focus on a \texttt{User Story} which is a \texttt{BO} that is used in planning, acting as a \textsl{Backlog Item}.
Other attributes include the \textsl{Level of Detail}, \textsl{Frequency of Changes}, and \textsl{Representation Format}.
In this example, we include extra explanatory text for the attributes in parentheses.
Two \texttt{MIs}, the \texttt{Development Team} and the \texttt{Product Management Team}, use this \texttt{BO} for coordination.
\texttt{Developers} and \texttt{Product Owner} roles are part of these \texttt{MIs}, respectively.
\texttt{Usage} for the \texttt{Developer} is captured via an association class, the attributes indicating that the \texttt{User Story} is easily accessible, critical, but with low stability, amongst other things.
A similar \texttt{Usage} class captures usage of the \texttt{BO} by the \texttt{Product Owner}.
The \texttt{Product Owner} is part of a \texttt{Forum of Product Owners} who make up the \texttt{Governance Team} for the \texttt{User Story BO}.
The \texttt{Governs} association class captures attributes of the governance process, e.g., they coordinate using the JIRA tool and meetings, and coordinate at least once per agile sprint.
Note that due to time restrictions, this model is incomplete, thus a blank value for some of the object attributes.
We consider how an instance model like this could be analyzed in the next section.

\subsection{BOMI Analysis}
\label{sec:analysis}
Although the process of creating a BOMI instance model is useful to understand BOs and MIs, one can go a step further and use the instance model created to detect potential issues or ``smells'' in the BOMI configuration, similar to the idea of smells in models or source code~\cite{van2002java,arendt2010uml}.
The idea is that these smells can be detected and discussed, determining if there is an underlying problem. This analysis and discussion would be conducted by those having a higher-level view of an organization, e.g., team leaders, project managers. The overall aim is to promote potential beneficial changes in the BOs, MIs, and ways of working. 

We can detect these smells within a BO, or across relationships in the model.
For example, we can detect smells within individual attributes:  low modularity, high maintainability, not up to date, not internally consistent, or not externally consistent.
We can also detect possible smells between attributes, including: having a high level of detail but a high frequency of change, meaning that frequent changes may be difficult and involve changing many elements; and being in an early lifecycle stage (planning) yet being very infrequently changed, or being in a later lifecycle change (deprecate, retire) yet having a high frequency of change.

\begin{table}
		\caption{Example smells in BOMI model instances with OCL expressions.} 
		\label{tab:smells}
		\centering
		\setlength{\tabcolsep}{3pt}
		\scalebox{0.9}{
			\begin{tabular}{@{}P{1.5cm}   P{3.4cm}  P{8.7cm}@{}}
				\toprule
				\textbf{Type}                                 & \textbf{Description}                           & \textbf{OCL Expression} \\ \midrule
				Within BO                          & Low modularity                                       & context \texttt{BoundaryObj} inv LowModularity: self.\textsl{Modularity} $=$ \textsl{Low}          \\
				& Not internally consistent     
				&  context \texttt{BoundaryObj} inv InternalInconsistency: self.\textsl{InternalConsistency} $=$ \textsl{Low}                       \\ 
				& High level of detail and frequent change  
				&     context \texttt{BoundaryObj} inv DetailedHighChange: self.\textsl{LevelofDetail} $=$ \textsl{High} and self.\textsl{FrequencyofChange} = \textsl{High}                   \\ 
				& Later lifecycle and frequent change                
				&   context \texttt{BoundaryObj} inv LateHighChanges: (self.\textsl{LifecycleStage} $=$ \textsl{Deprecate} or self.\textsl{LifecycleStage} $=$ \textsl{Retire}) and self.\textsl{FrequencyofChange} $=$ \textsl{High}                      \\ \midrule
				\multirow{2}{1.5cm}{Within Usage}                      & Not fit for purpose                                  &   context Usage inv NotFit: self.\textsl{FitForPurpose} $=$ \textsl{Low}                       \\ 
				& High criticality and low stability                 
				&    context Usage inv CriticalUnstable: self.\textsl{Criticality} $=$ \textsl{High} and self.\textsl{Stability} $=$ \textsl{Low}                     \\ \midrule
				\multirow{2}{1.5cm}{Missing Elements/ Relation-ships} & No governance team                                   &  context \texttt{BoundaryObj} inv Governed: self.\textsl{Governed}$\rightarrow$ size $>$ 0                       \\ 
				& No one responsible for BO                            
				&  context \texttt{BoundaryObj} inv Responsible: self.\textsl{Responsible}$\rightarrow$ size $>$ 0                        \\ 
				& No one can  update BO                 
				&   context \texttt{BoundaryObj} inv Updated: self.\textsl{Updates}$\rightarrow$ size $>$ 0                         \\ \midrule
				\multirow{2}{1.5cm}{Across Elements}                    & Governing roles should use BO                        & context \texttt{BoundaryObj} inv GovernsUses: self.\textsl{Governs}$\rightarrow$ forAll(g $\vert$ g.\textsl{PartOf}$\rightarrow$ select(r $\vert$ r.\textsl{Uses} = self)$\rightarrow$size $>$ 0)                       \\ 
				& High frequency of change but low frequency of coord
				&  context \texttt{BoundaryObj} inv GovernsUses: self.\textsl{FrequencyofChange} $=$ \textsl{High} and self.\textsl{Governs}$\rightarrow$ select(g $\vert$ g.\textsl{FrequencyofCoordination} $=$ \textsl{Low})$\rightarrow$size $>$ 0                        \\ 
				\bottomrule
		\end{tabular} }
\end{table}  
Similarly, with the \texttt{Usage} association class, smells include not being fit for purpose, or high criticality with low stability or low accessibility.
For instance, in Fig.~\ref{fig:Example1}, usage of the BO by both the developer and product owner is critical but the stability is low.
Is it acceptable for something so critical to change so frequently?  Looking into the BO, we see the lifecycle stage is planning, so the organization may argue that high criticality and low stability is unavoidable for key artifacts like user stories in this early stage.  If the artifact was instead in an operational stage, this situation may pose more of a problem.

We can also detect smells at a broader level, e.g., the BO has no governance team, or no one responsible for it.
Our company partners suggest that those governing a BO should also use it, to ensure that they are aware of how the BO is used.
It can also be checked whether there exists someone who can update and delete the BO.
And, if the \texttt{Usage} is critical, or if the frequency of change is high, the \texttt{Governs} class should likely have a high frequency of coordination.

We summarize how automatic checking of some of these smells could look using OCL expressions~\cite{cabot2012object} in Table~\ref{tab:smells}.
In our case, eventual tool support should allow the model to be drawn without necessarily following these expressions, capturing reality with smells.
These expressions could be checked after a first version of an instance model is created.
The output of such a check should be discussed within an organization, to determine if the smell is a problem in reality, and to discuss what sort of changes could be made.

\section{Evaluation}
\label{Evaluation}
The final step in our method was a 1.5-hour online workshop in April 2020 to try out the metamodel, method, and smell ideas with seven representatives from four companies, described in Table~\ref{tab:participants}.
The participants included systems engineers, requirements specialists, and tooling specialists.
During the workshop, we reserved 20 minutes for a review of BOMI concepts and to introduce the new metamodel using prepared material\footref{workshopmaterial}.
We then split off into four virtual break-out rooms for 30 minutes of modeling instance models in focused sessions.
Each room had at least one researcher and the representatives from one company.
The researchers went through the guiding method questions from Sec.~\ref{sec:method} and drew an instance model based on the answers of the participants, sharing their screen. 

Despite the short time-frame, we were able to get four relatively complete models (e.g., Fig.~\ref{fig:Example1}), with the statistics in terms of element type used shown in Table~\ref{tab:models}.
We opted to focus on one BO at a time; thus, each model had only one BO.
The modelers were also able to capture 2-5 MIs, 1-5 Usage association classes, 1-4 Drivers, and one Government Team and Governs association class per model.
Some of the attribute information for each model was filled in, but many attributes were left blank due to time restrictions.

\begin{table*} [b]		
		\setlength{\tabcolsep}{6pt}
		\caption{Element count of four instance models from the workshop.} 
		\label{tab:models}
		\centering	

			\begin{tabular}{@{}lllp{1cm}llp{2.9cm}p{1.4cm}@{}}
				\toprule
				\textbf{Model} & \textbf{BO} & \textbf{MI} & \textbf{Usage} & \textbf{Driver} & \textbf{Role} & \textbf{Governance~Team} & \textbf{Governs} \\ \midrule
				C1    & 1   & 2   & 2     & 1      & 2    & 1& 1       \\
				C2    & 1   & 3   & 1     & 1      & 5    & 1& 1       \\
				C3    & 1   & 5   & 5     & 4      & 0    & 1& 1       \\
				C4    & 1   & 3   & 1     & 2      & 2    & 1& 1       \\
				\bottomrule
		\end{tabular}
\end{table*}  

The final 30 minutes (allowing for short breaks) was used to discuss our experiences and gain feedback, with several of the authors taking notes.
The authors then met to share and review our notes, consolidating and discussing experiences.
Feedback included that the current typing hierarchy for MIs was often hard to apply, and MIs are often multi-dimensional.
To deal with this, we allowed MIs to have more than one type in the updated metamodel.
We also acknowledge that our current list of possible types (\textsl{MIType} in Fig.~\ref{fig:conceptualmodel}) may not be complete.
Previously, instead of the \texttt{Driver} class, we had an Ocean association class between \texttt{MI}s with a driver attribute.
We noted in our modeling exercises that \texttt{MI}s can have many drivers and can share drivers.
Thus, we reworked the Ocean association class to the current \texttt{Drivers} class.
We also made note that most of the attribute descriptions were hard to capture with enumerations (High/Medium/Low) and that we often needed free text descriptions to capture the subtleties, e.g., frequency of change varying depending on the lifecycle stage.
Finally, we made many small improvements to the class attributes.
We used all of this feedback to create the final version of the metamodel presented in Sec.~\ref{sec:findings}.
The previous three versions of the model can be found in our online appendix\footref{workshopmaterial}.

Our modeling sessions did not give us extensive time to apply the smell analysis examples as described in Sec.~\ref{sec:analysis}, and we were also hindered by the incompleteness of some of the instance model attributes.
However, we presented some draft smells and asked for feedback from the participants.
We generally asked ``Can the current issues with the BO be captured in the model?''
Although the participants were not opposed to automated checks as described in Sec.~\ref{sec:analysis}, they were more interested in human-centered manually-detected smells, e.g., ``Can I draw this?''
For them, the first and most important smell is whether the participants had the knowledge to instantiate the metamodel.
Our participants also suggested a smell having to do with the complexity of the overall model: ``I can draw it, but it is a mess'', indicating that the overall design of their BOMI situation could be overly complex and poorly thought-out.
Therefore, model complexity checks or basic checks such as for cohesion and coupling may be useful.
Our participants also suggested the check that those responsible for governance should also be users, and that the governance team should consist of a diverse set of roles or islands, i.e., not just be made up by one type of user.
Some of these smells could be expressed formally over the model, as in Sec.~\ref{sec:analysis}, but others can instead be included as points to consider in the methodology.

Overall, our company partners were positive about the experience.
Based on their interest, we are currently arranging longer sessions for two out of four companies, inviting further internal participants knowledgeable about key BOs.

\section{Related Work}
\label{sec:realted}
A number of related conceptual modeling approaches have been proposed.

\textbf{Knowledge Management.} Our work bears similarities to approaches that focus on modeling for knowledge management, e.g.,~\cite{Ale2014,strohmaier2007analyzing}.
Here the focus is often knowledge creation, distribution, representation, and retrieval.
Our approach captures some of these elements in the BOMI metamodel, including the format of the BO, its purpose, and users.
However, our focus is less about capturing implicit knowledge through a global strategy and more about understanding the way that diverse organizational islands coordinate knowledge through artifacts.

Other related work uses patterns to detect potential problems in information flows, e.g., consecutive transformations, which are similar to our notion of smells~\cite{schneider2009modeling}.
Our focus is less on the flow of information but more on effective coordination, thus our specific smells are quite different compared to~\cite{schneider2009modeling}.

\textbf{Agent-Orientation.} Our work bears some similarity to agent-oriented or multi-agent system modeling which emphasizes the rational behavior of individual agents in a system, e.g.,~\cite{gonccalves2019istar4rationalagents,jureta2005agent}.
Most of this work has an exchange of resources by agents through some form of dependency.
Although agent concepts could be used to capture MI, the islands are more like social groupings emerging due to various drivers, and often do not act together as a sentient and autonomous whole.
Similarly, BO could be resource dependencies, but our concept of BO is richer, and we place more emphasis on the means of use and attributes of BO, compared to resources in agent-oriented modeling. 

BOMI is in line with the Comakership organizational pattern~\cite{colombo2006multi}, with our notion of smells fitting with the idea of continuous improvement.
However, these patterns focus on inter-organizational coordination, while BOMI covers inter-team coordination, and BOMI does not make use of i* or intentions, with attributes such as ``Purpose'' in the BO fulfilling this role to a lesser degree. 

\textbf{Communication.} Work in~\cite{oliveira2007towards} introduces ontologies for collaboration, communication, and cooperation, with several elements and components echoed by our BOMI metamodel.
However, their focus is not on supporting diverse groups as with our MI, or on the attributes and specifics of the boundary objects or artifacts.
Some of the work which has focused on modeling communication focused on autonomous agents and their protocols, e.g.,~\cite{dignum1997communication}, while we focus on communication between MIs, always consisting of humans.

\textbf{Coordination.} Related work on coordination modeling focuses on coordination between information systems rather than human-oriented MIs~\cite{norrie1994coordination}.
In this view, coordination between systems can be captured via APIs, a type of BO.
Previously, benefits and limitations of languages for capturing APIs have been investigated~\cite{horkoff2018modeling}, e.g., i* and e\textsuperscript{3} value modeling.
Although the focus lay more on the use and value of APIs and less on coordination between methodologically diverse groups, BOMI may still be beneficial for API analysis.

Further work is more process-oriented.
~\cite{wieringa2008value} applies e\textsuperscript{3} value modeling, process modeling, and physical delivery modeling to support cross-organizational coordination.
ActivityFlow focuses on supporting incremental and flexible workflow definitions, allowing for workflow coordination between organizations~\cite{liu1997activity}.
BOMI takes a static, rather than process-oriented view, as our partner companies, with an agile mindset, focus less on workflows and more on practices.

\textbf{Ecosystems.}  Work in ecosystem modeling is also related (e.g.,~\cite{boucharas2009formalizing,yu2011understanding}), as our BOMI approach can be said to produce a type of ecosystem model; however, existing ecosystem models focus more on external coordination, where the internal methodologies of a partner are more opaque.
Our BOMI models tend to have a mix of internal and external MIs and BOs, often with a particular focus on supporting diversity in internal ways of working.
\vspace{-0.3cm}
\section{Discussion and Future Work}
\label{Discussion}
	\vspace{-0.2cm}
We have presented a conceptual model for BOMI, described how we instantiated it together with four large-scale systems development companies, and derived example smells over the instances that can be checked with OCL constraints.
Concretely, we have found that the BOMI model allowed us to create initial models with a rather low time effort (20 minutes of introduction of general concepts plus 30 minutes of modeling).
Our participants were positive about the outcome of the session and the initial models allowed us to test our list of initial smells.
We believe that the described findings are a good starting point to evaluate and tailor the BOMI metamodel further.
For instance, tooling, access, and security information could be added to the model, e.g., to facilitate security analysis concerning boundary objects.

In this paper, we focused on BOMI-specific smells. General UML smells, e.g., related to the use of names, attributes, or ``data clumps''~\cite{arendt2010uml}, might also be applicable to BOMI models, and 
are an interesting area for future work.

Moreover, we propose to investigate the creation and use of an expressive domain-specific visual language and tool support to capture BOMI models.
Currently, we rely on UML class diagrams due to the availability of general modeling tools and the existing familiarity with class diagrams.
However, there might be stakeholders (e.g., project managers, sales representatives) that are not familiar with class diagrams and could benefit from a domain-specific language.

Finally, we plan to build on these findings to help companies proactively address coordination issues and facilitate the management of boundary objects in practice.
Concretely, we aim to conceive a constructive method to continuously analyze the current situation with key stakeholders, propose actions for improvement, and mechanisms to assess the impact of implemented changes.

\textbf{Threats to Validity.} To improve \textit{internal validity/credibility}, we used an interactive modeling process with open questions, triangulated the experiences of the participating companies, and aimed to provide detailed descriptions in this paper.
A cross-company workshop was used to present the intermediate findings and perform member checking with the participants.

A threat to \textit{construct validity} relates to the nature of the domain we model.
The concepts of boundary objects and methodological islands can be misunderstood and interpreted in various ways.
We intended to provide clear definitions and engaged in a long-term project with the participating companies to ensure a common understanding of the concepts.

Considering \textit{external validity}, we used a sample of four large-scale companies that develop embedded systems.
We believe this sample provides valuable insights, but acknowledge we may have different findings with a different sampling  approach.
We describe the companies' characteristics in this paper to facilitate the assessment of what findings might be transferable to other contexts.

With respect to \textit{reliability}, the previously acquired knowledge of the participating companies in the project is a potential threat.
As stated before, we have previously collaborated on boundary objects and methodological islands, which will not be the case for other researchers or research contexts.
However, the general notation used in this paper is rather straight-forward and comprehensible for other modelers, which facilitates replication.
Moreover, we have made the explanatory material and models available online.
\vspace{-0.3cm}
\section{Conclusions}
\label{Conclusions}
\vspace{-0.2cm}
In this paper, we have focused on the challenge of inter-team coordination and knowledge management in large-scale systems development using diverse development practices.
While initial empirical studies existed, there has been a lack of systematic modeling approaches that can support practitioners in modeling their current and diverse coordination settings, and analyzing them to identify issues.
To address this issue, we proposed a conceptual model that can be used to model methodological islands (i.e., groups that work with a different methodology than their surrounding organization) and boundary objects between them (i.e., artifacts that can be used to create a common understanding across sites and support inter-team coordination).
We presented an initial list of bad smells that can be leveraged to detect issues and devise suitable strategies for inter-team coordination in large-scale development.
We evaluated the conceptual model together with four large industrial companies developing complex systems and present our positive evaluation results.

We plan to build onto these findings to devise a constructive method supporting the analysis of coordination issues and suggesting improvement strategies, as well as mechanisms to continuously assess the effect of these strategies. \newline

\textbf{Acknowledgments.} This work was partially supported by the Software Center Project 27 on Requirements Engineering for Large-Scale Agile System Development and the Wallenberg AI, Autonomous Systems and Software Program (WASP) funded by the Knut and Alice Wallenberg Foundation.

\end{document}